\documentclass[a4paper,11pt]{article}
\pdfoutput=1 % if your are submitting a pdflatex (i.e. if you have images in pdf, png or jpg format)

\usepackage{jinstpub} % for details on the use of the package, please see the JINST-author-manual

\title{Intensified Tpx3Cam, a fast data-driven optical camera with nanosecond timing resolution for single photon detection in quantum applications}

\author[a]{Andrei Nomerotski,\note{Corresponding author.}}
\author[a]{Matthew Chekhlov,}
\author[a]{Denis Dolzhenko,}
\author[b]{Rene Glazenborg,}
\author[a]{Brianna Farella,}
\author[a]{Michael Keach,}
\author[a]{Ryan Mahon,}
\author[b]{Dmitry Orlov,}
\author[c,d]{Peter Svihra}

\affiliation[a]{Brookhaven National Laboratory, Upton NY 11973, USA}
\affiliation[b]{Photonis Netherlands BV, 9301 ZR Roden, The Netherlands }
\affiliation[c]{Czech Technical University, 115 19 Prague, Czech Republic}
\affiliation[d]{CERN, 1211 Geneva 23, Switzerland}

\emailAdd{anomerotski@bnl.gov}

\abstract{We describe a fast data-driven optical camera, Tpx3Cam, with nanosecond scale timing resolution and 80 Mpixel/sec throughput. After the addition of intensifier, the camera is single photon sensitive with quantum efficiency determined primarily by the intensifier photocathode. The single photon performance of the camera was characterized with results on the gain, timing resolution and afterpulsing reported here. The intensified camera was successfully used for measurements in a variety of applications including quantum applications. As an example of such application, which requires simultaneous detection of multiple photons, we describe registration of photon pairs from the spontaneous parametric down-conversion source in a spectrometer. We measured the photon wavelength and timing with respective precisions of 0.15~nm and 3~ns, and also demonstrated that the two photons are anti-correlated in energy.
}


\begin{document}
\maketitle
\flushbottom

\section{Introduction}
\label{sec:intro}

The main motivations for fast imaging are studies of fast processes, measuring time of flight and looking for time coincidences of photons. The fast imaging approaches can be generally divided into two types: synchronous and asynchronous. In the former case, a synchronous signal, for example from a pulsed laser, is available, alerting that photons will be coming to the detector so some preparatory operations can be performed. In the latter case, which is more challenging, the measurement must be data driven meaning that a signal can be time stamped independently in each pixel. It is these kinds of cameras that are very attractive for applications where photons are arriving and need to be time stamped at any time in a continuous mode of operation.

A variety of sensing technologies are available for registration of single optical photons, which include approaches with external amplification:
iCCD and
iCMOS \cite{Brida2009, Reichert2017, Jost1998, Jachura2015, Fickler2013}; with internal amplification:
EMCCDs \cite{Zhang2009, Avella2016, Moreau2019} and
SPADs \cite{Gasparini2017, Perenzoni2016, Lee2018, Morimoto2020, Lubin2021}; and superconductive technologies:
SNSPD \cite{Divochiy2008, Zhu2020, Korzh2020} and TES \cite{Cabrera1998, Lita2008}; see comprehensive reviews of the subject in \cite{Hadfield2009, Seitz2011, SinglePhoton2013ii}.
The camera described here, Tpx3Cam, is an intensified hybrid CMOS imager, which is capable of time stamping photon flashes by adding an optical sensor to the Timepix3 readout chip. Single photon sensitivity is provided by an external optical intensifier. 

Below we describe different aspects of the camera performance, which are especially important for quantum applications. Firstly, in Section \ref{sec:camera} we provide details on the intensified camera. Section \ref{sec:characterization} discusses several studies important for the single photon detection and Section \ref{sec:spectrometer} describes a fast spectrometer used for characterization of a quantum single photon source.

\section{Tpx3Cam fast camera}
\label{sec:camera}

Imaging of single photons in the experiments mentioned here was performed with a  time-stamping camera, Tpx3Cam \cite{timepixcam, tpx3cam, Nomerotski2019, ASI}. The camera has a silicon optical sensor with high quantum efficiency (QE) \cite{Nomerotski2017}, which is bump-bonded to Timepix3 \cite{timepix3}, an application specific integrated circuit (ASIC) with $256 \times 256$ pixels of $55 \times 55$ $\mu$m$^2$. The electronics in each pixel processes the incoming signals to measure their time of arrival (TOA) for hits that cross a predefined threshold, around 600 electrons, with 1.56~ns precision. The information about time-over-threshold (TOT), which is related to the deposited energy in each pixel, is stored together with TOA as time codes.
The Timepix3 readout is data driven with pixel deadtime of only 475~ns~+~TOT, allowing for multi-hit functionality for each pixel, independently from the other ones, and fast, 80 Mpix/sec, bandwidth \cite{heijdenSPIDR}.

\subsection{Intensified Camera}

\begin{figure*}[!htb]
    \centering
    \includegraphics[width=0.543\linewidth]{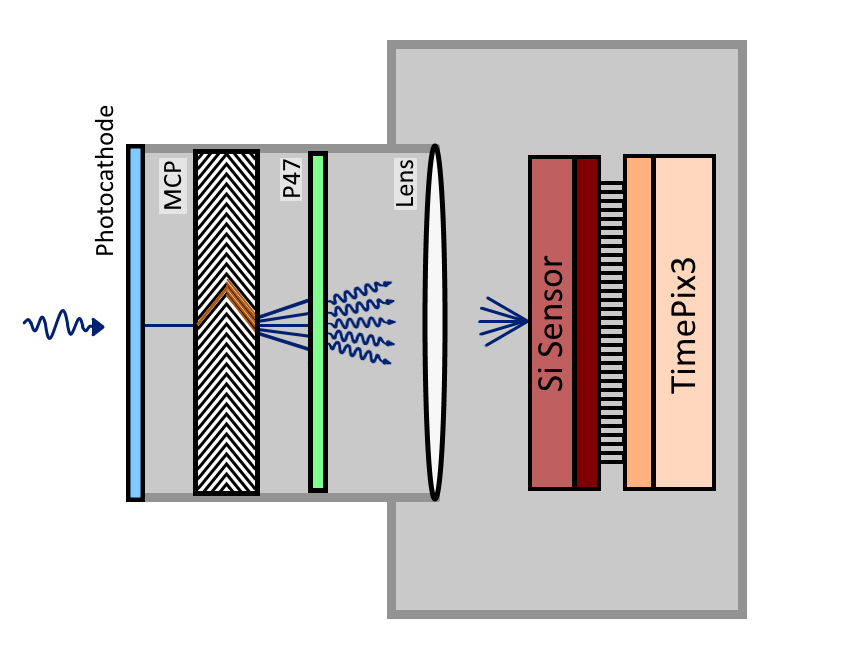}
    \includegraphics[width=0.45\linewidth]{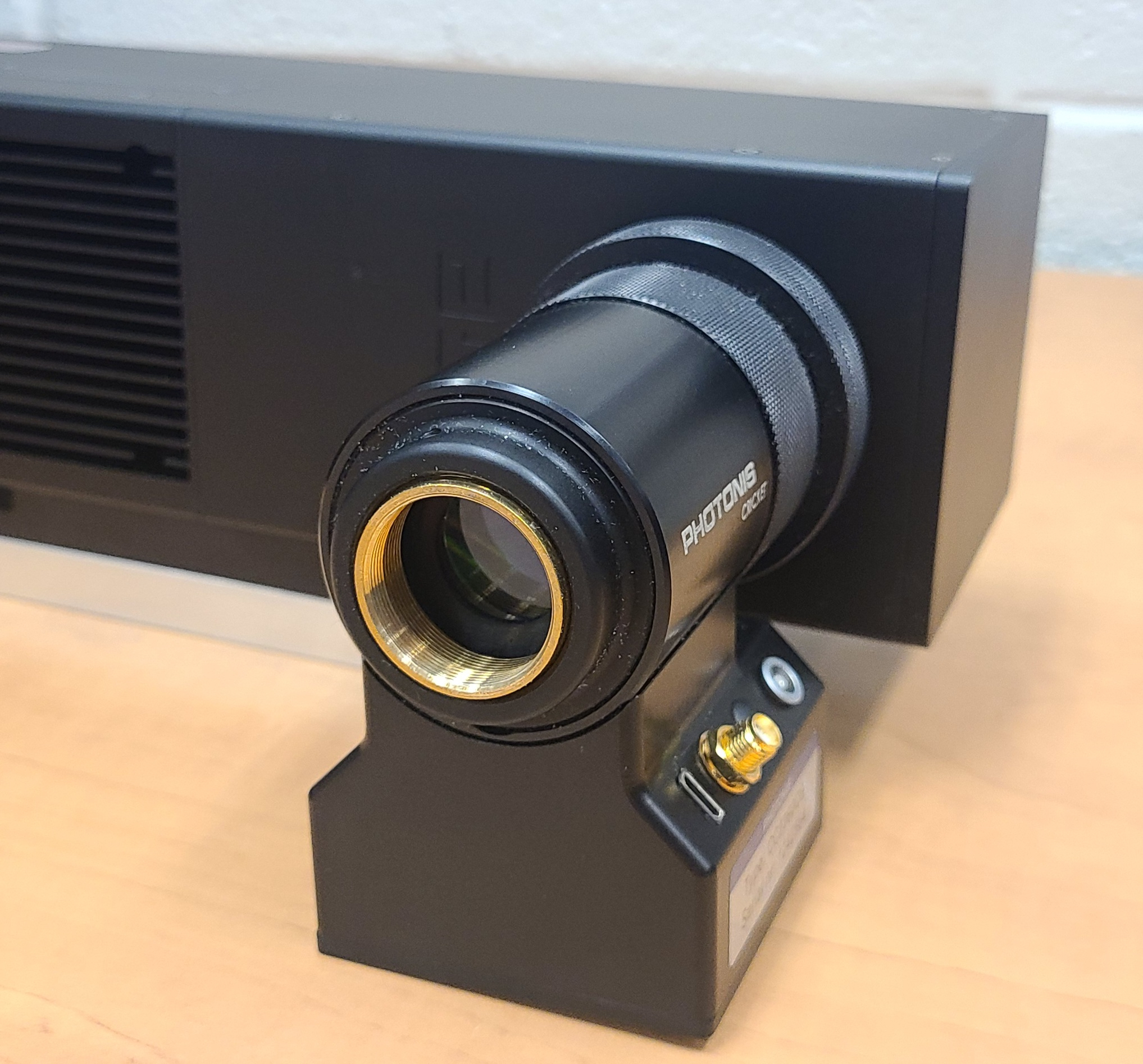}
    \caption{Left: Schematic layout of the intensified Tpx3Cam camera, not to scale. Right: photograph of the camera with Cricket$^{\rm{TM}}$.}
    \label{fig:photo}
\end{figure*}

For the single photon sensitive operation, the signal is amplified with addition of Cricket$^{\rm{TM}}$ \cite{Photonis} with integrated image intensifier, power supply and relay optics to project the light flashes from the intensifier output window directly on to the optical sensor in the camera. The image intensifier is a vacuum device comprised of a photocathode followed with a micro-channel plate (MCP) and fast scintillator P47 with risetime of about 7~ns and maximum emission at 430 nm \cite{Winter2014}. The quantum efficiency of the camera is determined primarily by the intensifier photocathode with a variety of photocathodes available \cite{Photonis, Nomerotski2019}. As an example, the hi-QE-green photocathode has QE of about 30\% in the range of 400 - 480~nm \cite{Orlov2016}. The MCP in the intensifier had an improved detection efficiency close to 100\% \cite{Orlov2018}.
Figure \ref{fig:photo} shows schematically the layout of the intensifier and camera, and their photograph. 

Similar configurations of the intensified Tpx3Cam were used before for characterization of quantum networks \cite{Ianzano2020, Nomerotski2020}, quantum target detection \cite{Yingwen2020, Svihra2020}, single photon counting and ray tracing \cite{sensors2020, Zhang2021, Gao2022, Zhang2022}, studies of micromotion in ion traps \cite{Zhukas2021_1, Zhukas2021,Kato2022}, neutron detection \cite{DAmen2021,losko2021, Yang2021}, optical readout of time-projection chamber (TPC) \cite{Roberts2019} and lifetime imaging \cite{Hirvonen2017, Sen2020, Sen2020_1} studies. 
 The Timepix based single photon sensitive cameras with direct registration of MCP electrons with the Timepix ASIC metal pads have also been produced in the past \cite{valerga2014}.

\subsection{Post-processing}

\begin{figure*}[!htb]
    \centering
    \includegraphics[width=0.8\linewidth]{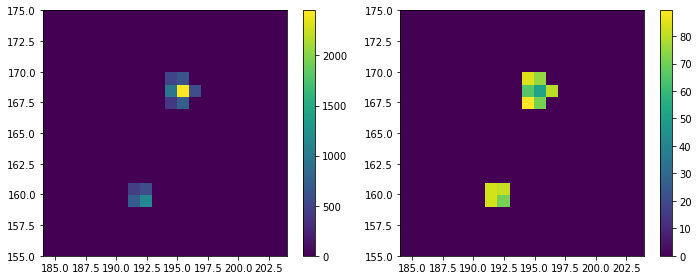}
    \caption{Raw TOT (left) and TOA (right) values in nanoseconds for pairs of single photon hits in Tpx3Cam.}
    \label{fig:hits}
\end{figure*}

Figure \ref{fig:hits} shows an example of raw TOT and TOA information for a pair of single photons from a spontaneous parametric down-conversion (SPDC) source registered in Tpx3Cam. As one can see, the photons appear as small groups of hit pixels, so
after ordering in time, the pixels are grouped into the "clusters" using a recursive algorithm \cite{tpx3cam} within a predefined time window, 300~ns. 
Since all hit pixels measure TOA and TOT independently and provide the position information, it can be used for centroiding to determine the coordinates of single photons. The TOT information is used for the weighted average, giving an estimate of the x, y coordinates for the incoming single photon with improved spatial resolution \cite{Hirvonen2017}. The timing of the photon is estimated by using TOA of the pixel with the largest TOT in the cluster. 
%The above ToA is then adjusted for the so-called time-walk, an effect caused by the variable pixel electronics time response, which depends on the amplitude of the input signal \cite{Turecek_2016, tpx3cam}. With this correction a 2~ns time resolution (rms) can be achieved for single photons \cite{Ianzano2020}.

\section{Characterization of single photon performance}
\label{sec:characterization}

\subsection{Intensifier gain and signal amplitude}

\begin{figure*}[!htb]
    \centering
    \includegraphics[width=0.45\linewidth]{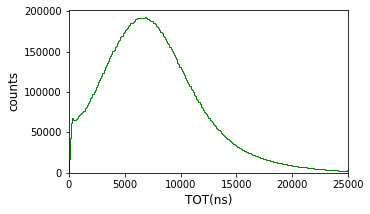}
    \includegraphics[width=0.45\linewidth]{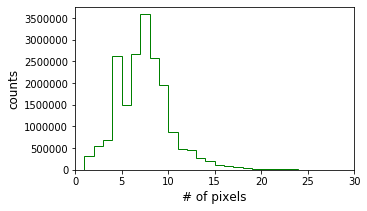}
    \caption{Left: time-over-threshold (TOT) distribution. Right: distribution of number of pixels in cluster.}
    \label{fig:TOT}
\end{figure*}

The left part of Figure \ref{fig:TOT} gives an example of the TOT distribution for the TOT sum over all pixels in the cluster. The distribution is broad due to large fluctuations of the MCP gain, so cannot be used for the photon counting. In this particular case the MCP gain is at maximum of about $10^6$ and the TOT distribution peaks at approximately 60,000 electrons. Accounting for the sensor quantum efficiency for the emission spectrum of P47, that means the P47 light flash had about 70,000 photons \cite{Nomerotski2017}. 
We can also see that the distribution for small TOT values has a sharp cutoff near zero due to the Timepix3 threshold. These measurements were performed for the hi-QE-red photocathode \cite{Photonis}. 

For the measurements described here and also in Sections 3.3 and 3.4, we used the dark counts of a hi-QE-red intensifier. The total dark count rate was about 80~kHz distributed uniformly across the photocathode. This rate is entirely due to the thermal photoelectrons from the photocathode. After going through the detection chain (MCP-P47-Tpx3Cam),
these signals look exactly the same and have same properties as signals from the single photons \cite{Nomerotski2017}.

The right part of Figure \ref{fig:TOT} shows the number of pixels in clusters. The maximum of the distribution is at seven pixels. The sub-peak at four pixels is explained by the symmetry of  $2\times2$ pixels square shape.

\subsection{Timing resolution}

\begin{figure*}[!htb]
    \centering
    \includegraphics[width=0.58\linewidth]{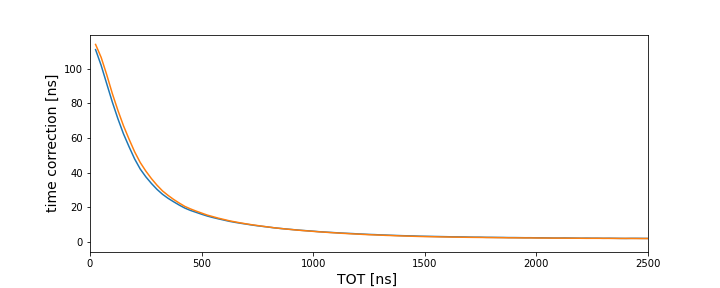}
    \includegraphics[width=0.41\linewidth]{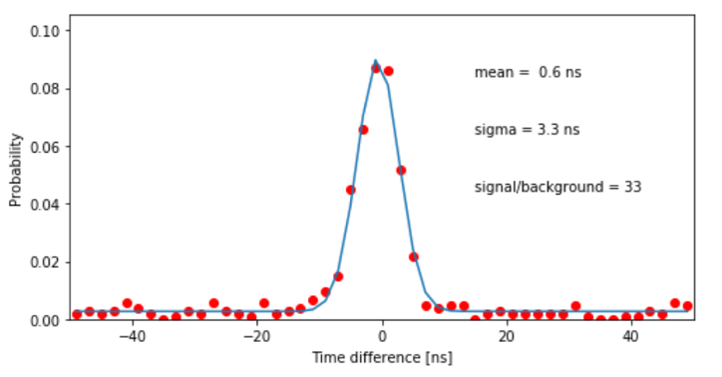}
    \caption{Left: Two instances of TOT correction for the same camera taken at different times; Right: Time difference distribution for a pair of simultaneous photons from SPDC source after the TOT correction.}
    \label{fig:timing}
\end{figure*}

Timing resolution is a critical parameter for many applications. The response of the linear discriminator used in Timepix3 has a time lag, which is dependent on the signal amplitude and can be as large as 100~ns for small signals near the threshold. To account for it and achieve the best possible resolution, a TOT correction must be applied \cite{Zhao2017, Tsigaridas2019}. The left part of Figure \ref{fig:timing} shows two instances of TOT correction for the same camera taken at different times as a function of TOT value. These measurements were performed with the same hi-QE-red intensifier as in Section 3.1.

The right part of Figure \ref{fig:timing} shows the time difference distribution for pairs of  photons from a SPDC source, which produce simultaneous pairs of photons, after the TOT correction. The time resolution (rms) per photon, assuming equal contributions for two photons in the pair, is $3.3/\sqrt{2}=2.4$~ns, a typical value achievable after the correction. It was shown that the Timepix3 timing resolution can be further improved by calibration of time offsets for individual pixels and time centroiding algorithms \cite{Heijhoff2021}.

\subsection{Afterpulsing}

\begin{figure}[!htb]
\begin{center}
\includegraphics[width=0.52\linewidth]{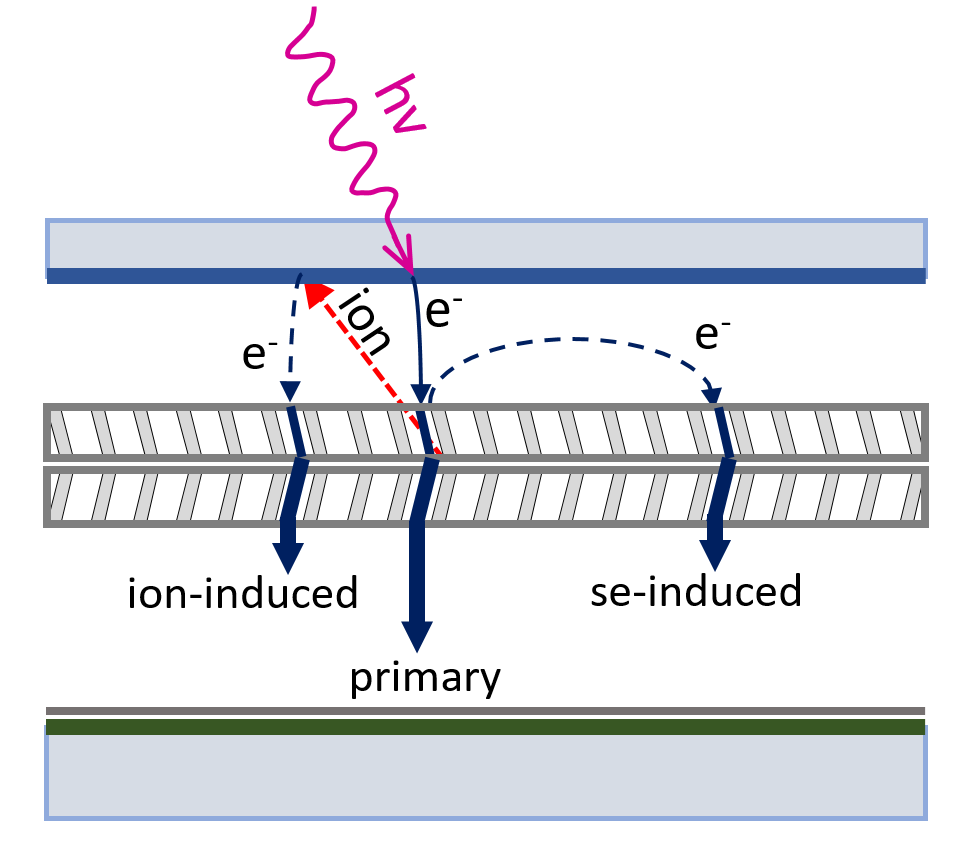}
\includegraphics[width=0.47\linewidth]{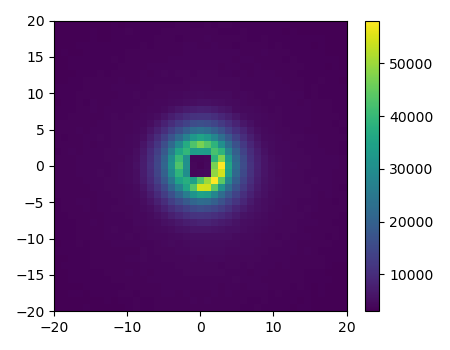}
\caption{Left: Two mechanisms of afterpulsing due to the ion feedback and secondary electrons. 
Right: Spatial distribution of the afterpulses. The empty square $3\times3$ pixels in the middle is due to the requirement of the two hits to be separated by at least one pixel.}
\label{fig:afterpulses}
\end{center}
\end{figure}

Afterpulsing is an important feature of the intensifier. It can be caused by secondary electrons or ions, which start another multiplication process in the vicinity of the primary pulse with a short delay \cite{Orlov2019}. The left part of Figure \ref{fig:afterpulses} illustrates these two afterpulsing mechanisms.

To find the afterpulses in the data, we look at the differences in time and position between the successive hits.
The right part of Figure \ref{fig:afterpulses} shows a map of position differences in $x$ and $y$ for all consecutive in time pulses. There is an obvious excess of pulses spatially close to each other due to the afterpulsing. It also has an obvious asymmetry of the azimuthal distribution, which indicates a preferred direction, likely related to the orientation of the capillaries in the MCP glass. We verified this by rotating the intensifier by 90$^\circ$ with respect to the sensor, observing that the asymmetry followed the rotation.

\begin{figure}[!htb]
\begin{center}
\includegraphics[width=0.4972\linewidth]{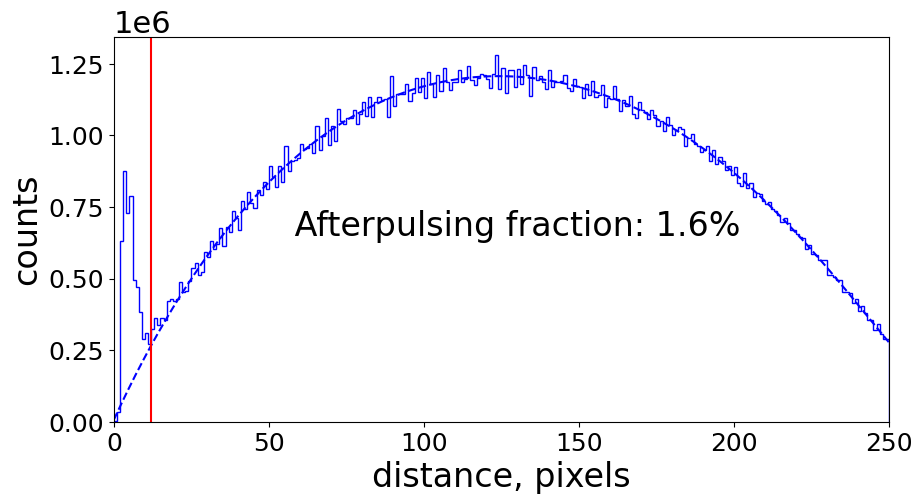}
\includegraphics[width=0.4972\linewidth]{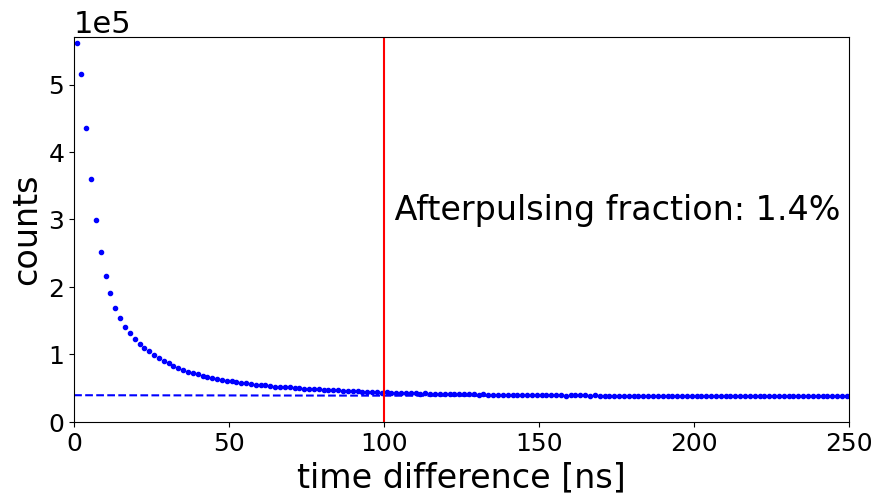}
    \caption{Left: Distribution of distances for two successive hits. Right: Distribution of time differences for two successive hits.}
    \label{fig:dist_fraction}
\end{center}
\end{figure}

To further characterize the afterpulsing, we plot the distributions of distance between the successive hits and their separation in time, see  Figure \ref{fig:dist_fraction}. Note that the flat background in the time difference distribution is explained by random coincidences of dark count hits, with total rate of about 80~kHz. There is an obvious excess for the small values in both distributions, which we estimate as deviations from the expected smooth behaviour and interpret as afterpulses. We found that the fraction of afterpulsing events closer than 10 pixels is 1.6\% and fraction of afterpulsing events within 100~ns is 1.4\%, which are satisfactorily close to each other values.

\subsection{Bias voltage dependence}

\begin{figure*}[!htb]
    \centering
    \includegraphics[width=1\linewidth]{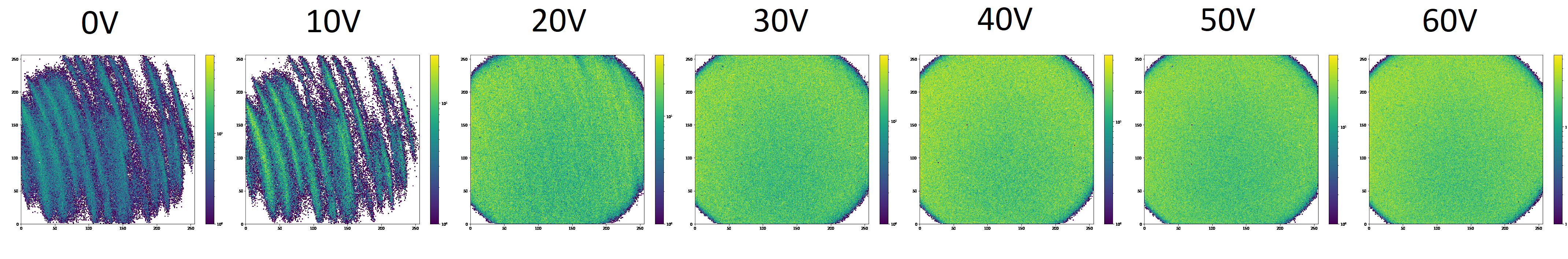}
    \caption{Normalized 2D occupancy for the intensifier dark counts for the silicon bias voltage between 0V and 60V.}
    \label{fig:OldCameraDCRNormed}
\end{figure*}

Fully depleting silicon sensor is important for the camera operation with high detection efficiency. Figure \ref{fig:OldCameraDCRNormed} shows the normalized dark count images of an intensifier for the silicon bias voltage between 0V and 60V in Tpx3Cam. One can see that for low values of the bias voltage the silicon is not depleted showing characteristic stripes, so-called tree rings, due to variations of its resistivity \cite{Holland2014, Park2017}. For the bias voltage above 30V, the structures are not visible, which agrees with expectations that the sensors are fully depleted at this voltage. A typical bias voltage used in the measurements is 50V.

\section{Characterization of fast spectrometer based on the camera}
\label{sec:spectrometer}

Multiple quantum applications of the camera have been already mentioned in Section \ref{sec:camera}. Key to its successful employment was the efficient detection of multiple photons in the same device, which allowed a straightforward identification of their coincidences. Below we provide a detailed description of one of those experiments involving a single photon spectrometer.

Spectral binning of single photons for two-photon interferometry is a promising venue for various quantum applications \cite{harvard2, Yingwen2020, Stankus20, Nomerotski20, Brown2022}. We implemented a simple dual spectrometer by sending two beams of collimated light on to a diffraction grating and then focusing them on to the intensified Tpx3Cam. The spectrometer was characterized using the argon spectrum, which has a large number of narrow lines.
%Our spectrometer allows analysis of multiple beams, which can be diffracted independently on to different areas of the camera.  
The resulting pattern is detected by the camera to produce an image like shown in the left part of Figure \ref{fig:argon}, with two horizontal stripes corresponding to two diffraction patterns produced by the same thermal argon source after splitting the beam in to two parts. In each of these stripes, distinct peaks can be seen, which 
can be interpreted as spectral lines of argon. The shown spectrum range is about 60~nm starting from 860~nm determined by the lens after the grating.

\subsection{Linearity and spectral resolution}

\begin{figure}[!htb]
\begin{center}
\includegraphics[width=0.35\linewidth]{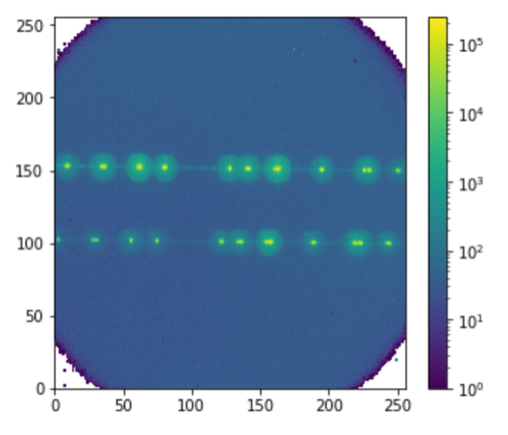}
\includegraphics[width=0.60\linewidth]{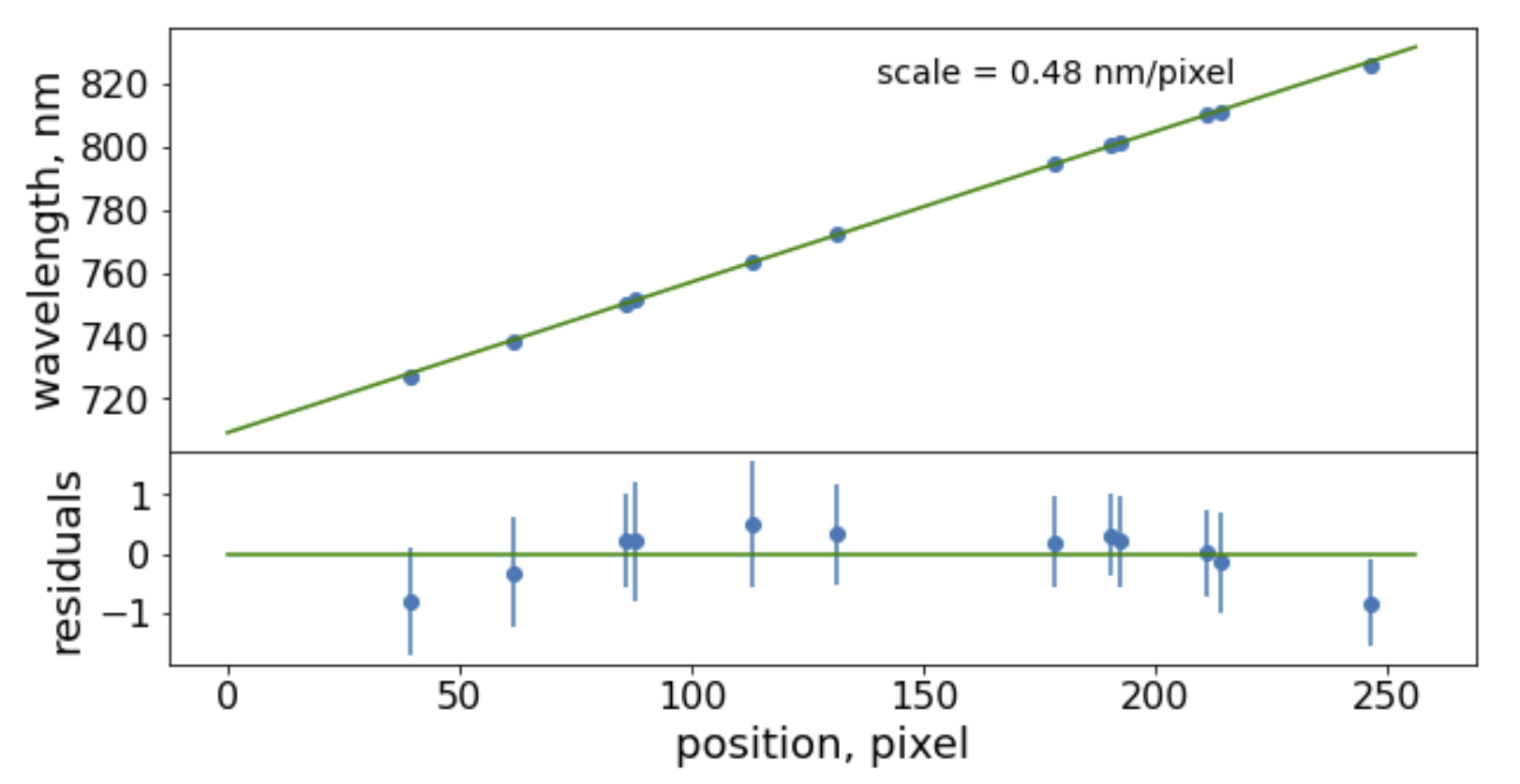}
    \caption{Left: Two argon spectra for two beams derived from the same argon lamp. Right: Spectrometer linearity, scale and residuals.}
    \label{fig:argon}
\end{center}
\end{figure}

Since positions of the lines are well known we can use them for calibrations to determine the conversion scale and linearity. The right part of Figure \ref{fig:argon} provides the scale and linearity graph with corresponding residuals. 

%All residuals in the linearity plot were within one spectral resolution of the fit line, which indicates that our measurements and methods were accurate within experimental limitations and the linearity was very good.

%We measured spectral resolution and scale for various images with focal lengths of 100~mm, 200~mm, and 400~mm. As expected, scale changed linearly with respect to focal length. 

\begin{figure}[!htb]
\begin{center}
\includegraphics[width=0.60\linewidth]{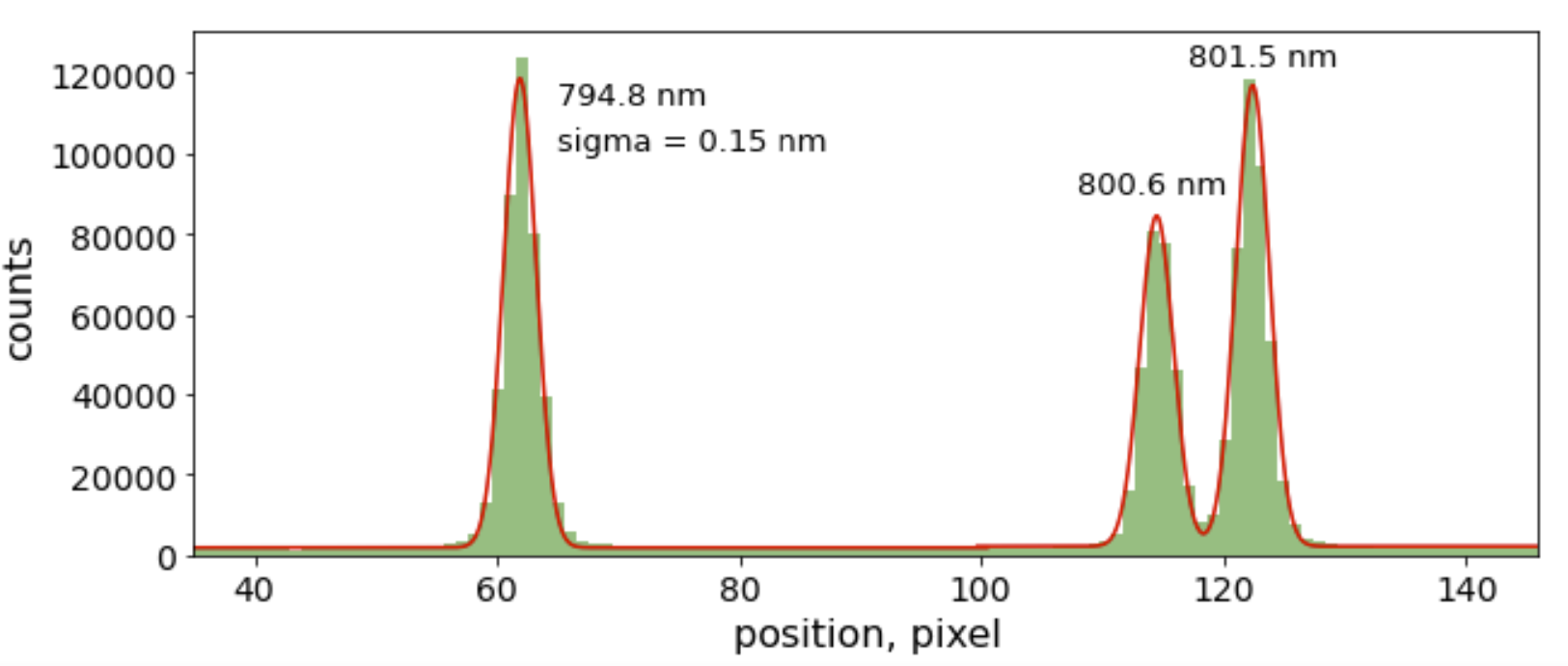}
\includegraphics[width=0.35\linewidth]{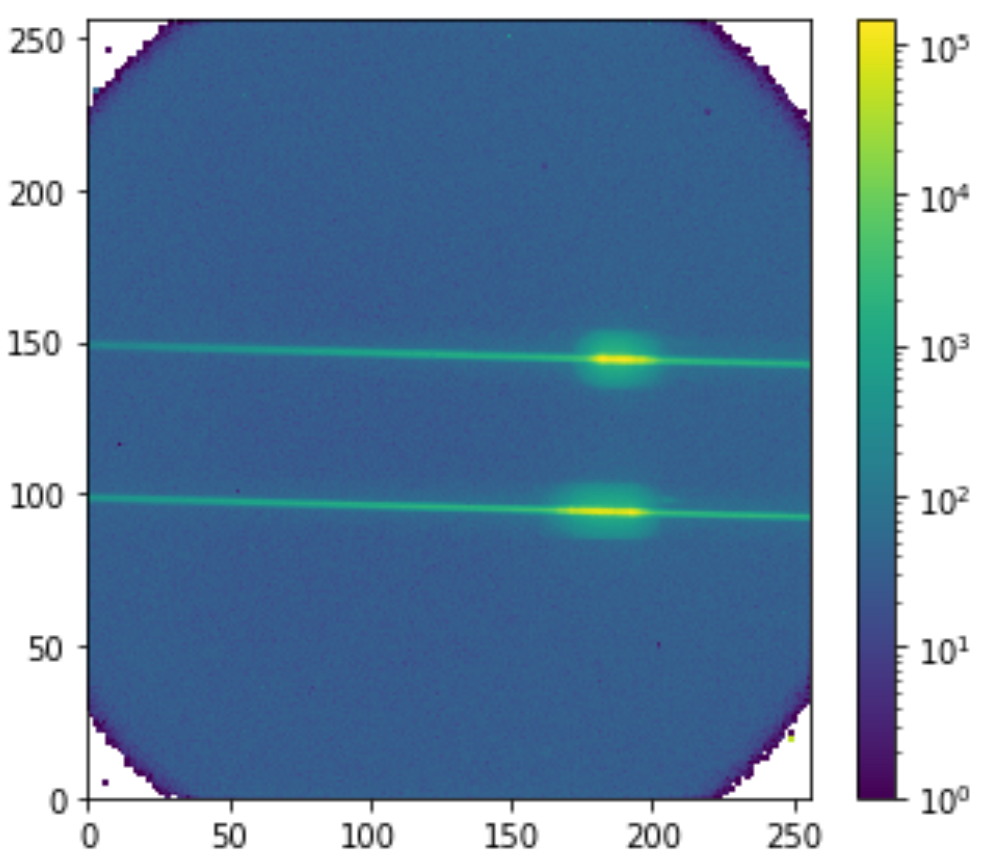}
    \caption{Left: Three argon lines in spectrometer with increased magnification. Right: Signal (top stripe) and idler (bottom stripe) photons in the spectrometer.}
    \label{fig:3lines}
\end{center}
\end{figure}

Figure \ref{fig:3lines} shows three spectral lines: 794.8, 800.6 and 801.5~nm, for the same spectrometer with increased magnification. %compared to the spectrum shown in Figure \ref{fig:argon}.
The lines are fitted with a Gaussian function with sigma 0.15~nm.

\subsection{Spectral analysis of SPDC source}

\begin{figure}[!htb]
\begin{center}
\includegraphics[width=0.4\linewidth]{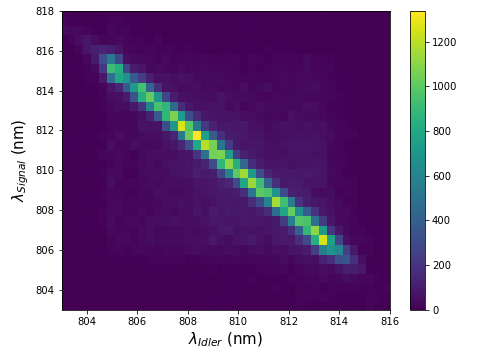}
\includegraphics[width=0.41\linewidth]{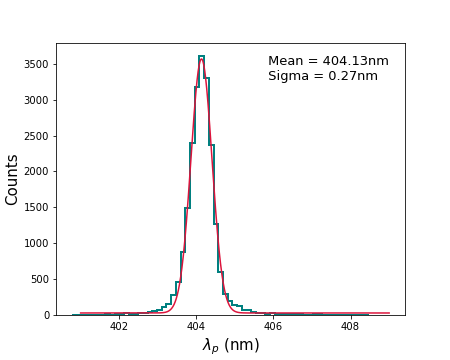}
    \caption{Left: Anti-correlation of photon wavelengths for the SPDC source. Right: Reconstructed wavelength of the pump photon using the wavelengths of the signal and idler photons, and argon spectrum calibration.}
    \label{fig:pump}
\end{center}
\end{figure}

To demonstrate the performance of the spectrometer for single photons we used a Thorlabs SPDC source of photon pairs \cite{Thorlabs}, where the signal and idler photons from the source were characterized in the two spectrometer channels. The full spectra of the photons are shown in the right part of Figure \ref{fig:3lines}. The signal and idler photons in the pair anti-correlate in the wavelength to conserve the energy of pump photon.
The left part of Figure \ref{fig:pump} shows dependence of the signal wavelength on the idler wavelength for photon pairs within a 20~ns time window, where this anti-correlation is apparent. The right part of Figure \ref{fig:pump} shows the reconstructed wavelength of the pump photon using the wavelengths of the signal and idler photons, and the argon spectrum calibration. 

%We note that the width of the distribution is wider than the width of argon lines above due to contributions of two independent wavelength measurements.

\section{Conclusions}

We described a fast optical camera with nanosecond scale timing resolution. The intensified version of the camera was successfully used in a variety of quantum applications, which required simultaneous detection and time stamping of multiple single photons. 

Possible improvement to the described approach would be to switch to the next generation readout chip, Timepix4 \cite{Llopart2022}, with timing resolution of 200~ps. The testing preparations are in progress though it remains to be seen if the optical path through the scintillator flashes registered by the silicon photodiode and Timepix4 front-end electronics would be fast enough to fully support the improved time resolution of Timepix4. 

%As an example of application, which require simultaneous detection of multiple photons, we detected photon pairs from the SPDC source in a spectrometer. We measured their wavelength with precision of 0.15\%  and demonstrated that the two photons are anti-correlated in energy.

\acknowledgments % equivalent to \section*{ACKNOWLEDGMENTS}       
 
This work was supported by the U.S. Department of Energy QuantISED award and BNL LDRD grant 22-22. M.C., B.F. and R.M. acknowledge support under the Science Undergraduate Laboratory Internships (SULI) Program by the U.S. Department of Energy.

% Produces the bibliography via BibTeX.

\bibliographystyle{unsrt}
\bibliography{IWORID2022}

\end{document}